\documentclass[12pt,epsf]{iopart}
\usepackage{epsfig}

\def\ra{\rangle}
\def\la{\langle}
\def\bege{\begin{equation}}
\def\ende{\end{equation}}
\def\begarr{\begin{eqnarray}}
\def\endarr{\end{eqnarray}}
\def\no{\noindent}
\def\non{\nonumber}

\def\ad{{{\hat a}^\dagger}}
\def\bd{{{\hat b}^\dagger}}
\def\cd{{{\hat c}^\dagger}}
\def\dd{{{\hat d}^\dagger}}

\def\ha{{\hat{a}}}
\def\hb{{\hat{b}}}
\def\hc{{\hat{c}}}
\def\hd{{\hat{d}}}

\begin{document}

\title[From Linear Optical Quantum Computing ...]  
{From Linear Optical Quantum Computing to Heisenberg-Limited Interferometry}

\author{Hwang Lee\dag, Pieter Kok\dag\ddag,  Colin P.\ Williams\dag,
and Jonathan P.\ Dowling\dag  
\footnote[3]{To
whom correspondence should be addressed (jonathan.p.dowling@jpl.nasa.gov)}
}

\address{\dag\ 
Quantum Computing Technologies Group, 
Section 367, Jet Propulsion Laboratory, \\
California Institute of Technology, MS 126-347,
 4800 Oak Grove Drive, CA~91109, USA}

\address{\ddag\ 
Hewlett Packard Laboratories, Bristol  BS34 8QZ, UK}

\begin{abstract}
The working principles of linear optical quantum computing are
based on photodetection, namely, projective measurements.
The use of photodetection
can provide efficient nonlinear interactions between photons
at the single-photon level,
which is technically problematic otherwise.
We report an application of such a technique to prepare 
quantum correlations as an important resource 
for Heisenberg-limited optical interferometry, 
where the sensitivity of phase measurements can be improved 
beyond the usual shot-noise limit.
Furthermore, 
using such nonlinearities,
optical quantum nondemolition measurements 
can now be carried out easily at the single-photon level.

\end{abstract}



\maketitle

\section{Effective nonlinearities from projective measurements}

Looking back, scalable quantum computation with linear optics was 
considered to be impossible due to the lack of efficient two-qubit logic gates, 
despite the ease of implementation of one-qubit gates.
Two-qubit gates necessarily need a nonlinear interaction
between the two photons, 
and the efficiency of this
nonlinear interaction is typically very tiny 
in bulk materials \cite{boyd91}.
However, Knill, Laflamme, and Milburn recently showed that
this barrier can be circumvented with effective nonlinearities
produced by projective measurements \cite{klm01}, 
and with this work 
scalable linear optical quantum computation (LOQC) becomes a reality.

Let us consider the Kerr nonlinearity, which can be described by
a Hamiltonian \cite{scully97}
\begarr
{\cal H}_{\rm Kerr} = \hbar \kappa \ad\ha \bd\hb,
\endarr
\no
where $\kappa$ is a coupling constant depending on the third-order
nonlinear susceptibility, and 
$\ad$, $\bd$ and $\ha$, $\hb$ are the creation and annihilation 
operators for two optical modes.
One convenient choice of the logical qubit can then be
represented by the two modes containing a single photon,
denoted as
\begarr
|0\ra_L &=& |0\ra_l ~|1\ra_k \non \\
|1\ra_L &=& |1\ra_l ~|0\ra_k ,
\endarr
\no
where 
$l,k$ represent the relevant modes, and
we have used the notation
$|\cdot\ra_L$ for a logical qubit, in order to
distinguish it from the photon-number states $|\cdot\ra_k$.

For a two-qubit gate, let us assign mode 1,2 for the control
qubit, and 3,4 for the target qubit.
Suppose now only the modes 2,4 are coupled under the 
interaction given by Eq.(1).
For a given interaction time $\tau$, the transformation
can be written as
\begarr
|0\ra_L |0\ra_L &&\rightarrow ~~|0\ra_L |0\ra_L 
\non \\
|0\ra_L |1\ra_L &&\rightarrow ~~|0\ra_L |1\ra_L 
\non \\
|1\ra_L |0\ra_L &&\rightarrow ~~|1\ra_L |0\ra_L 
\non \\
|1\ra_L |1\ra_L &&\rightarrow e^{i \varphi} |1\ra_L |1\ra_L ,
\endarr
\no
where $\varphi \equiv \kappa n_a n_b \tau$
and 
$n_a =\la \ad\ha \ra, n_b=\la \bd\hb\ra$. 
This operation yields a conditional phase shift \cite{turchette95}.
When $\varphi =\pi$, we have the two two-qubit gate
called the conditional sign-flip gate.
A typical two-qubit gate, controlled-NOT (CNOT), is then simply
constructed by using the conditional sign flip and
two one-qubit gates (e.g., Hadamard on the target, followed by
the conditional sign flip
and another Hadamard on the target).
In order to have $\varphi \sim \pi$ at the single-photon level,
however,  a huge third-order nonlinear coupling 
is required \cite{milburn89}.
Instead, Knill, Laflamme, and Milburn devised
a nondeterministic conditional sign flip gate
using nonlinear sign gate defined by
\begarr
\alpha|0\ra + \beta |1\ra + \gamma |2\ra
~\longrightarrow~ 
\alpha|0\ra + \beta |1\ra - \gamma |2\ra.
\endarr
The nonlinear sign gate can be implemented non-deterministically
by three beam splitters, two photo-detectors,
and one ancilla photon \cite{ralph01} (see Fig.\ 1).
   \begin{figure}
   \begin{center}
  \begin{tabular}{c}
   \includegraphics[height=3cm]{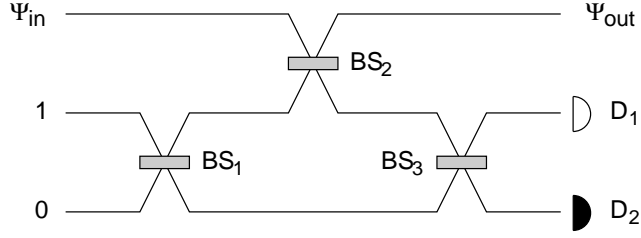}
   \end{tabular}
   \caption  
{A diagram for the nonlinear sign gate.
Conditioned upon a specific detector outcome,
the desired output state can be obtained 
by choosing appropriate transmission coefficients
of the beam splitters.
The success probability of the gate operation is 1/4,
but we always know when it succeeds.
}
\end{center}
\end{figure}
\no
The implementation of conditional sign flip gate is then
made by the combination
of the nonlinear sign gate and the physics of
Hong-Ou-Mandel (HOM) interferometer \cite{hom87}.
For two arbitrary qubits
\begarr
|Q_1\ra &=& \alpha_0 |0\ra_{L} + \alpha_1 |1\ra_{L}
= \alpha_0 |0\ra_1 |1\ra_2 + \alpha_1 |1\ra_1 |0\ra_2 ,
\non \\
|Q_2\ra &=& \alpha_0^\prime |0\ra_{L} + \alpha_1^\prime
|1\ra_{L}
=
\alpha_0^\prime |0\ra_3 |0\ra_4 + \alpha_1^\prime |1\ra_3 |0\ra_4 ,
\endarr

\no
the transformation of applying a condition sign flip gate can 
be written as
\begarr
|Q_1\ra |Q_2\ra
&\Rightarrow&
\alpha_0 \alpha_0^\prime |0\ra_{L} |0\ra_L 
+ \alpha_0 \alpha_1^\prime |0\ra_L |1\ra_{L}
+ \alpha_1 \alpha_0^\prime |1\ra_L |0\ra_L 
- \alpha_1 \alpha_1^\prime |1\ra_L |1\ra_L 
\non \\
&&=
\alpha_0 \alpha_0^\prime |0,1,0,1\ra
+ \alpha_0 \alpha_1^\prime |0,1,1,0\ra
+ \alpha_1 \alpha_0^\prime |1,0,0,1\ra
- \non \\
&& \qquad \alpha_1 \alpha_1^\prime |1,0,1,0\ra  .
\endarr

\no
where
the modes 1 and 2 are designated for the control qubit,
and 3 and 4 are for the target qubit. 
A sign change happens
only when there is one photon in mode 1
and one photon in mode 3.

  \begin{figure}[b]
   \begin{center}
   \begin{tabular}{c}
   \includegraphics[height=3.5cm]{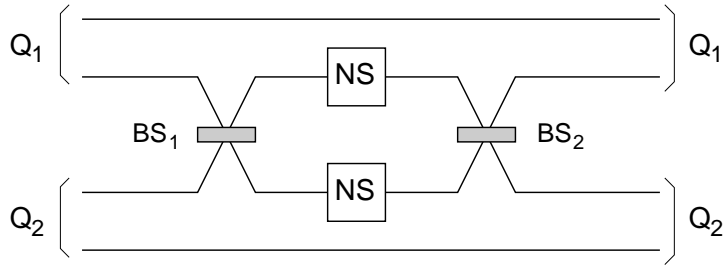}
   \end{tabular}
   
   \caption[example] 
{Nondeterministic conditional sign-flip gate.
The relevant optical modes are assigned as \{2,1,3,4\} from the top.
When the modes 1, and 3 contain one photon each, 
$|1,1\ra_{1,3}$ ($|1\ra_L|1\ra_L$), 
it becomes $|2,0\ra_{1,3}-|0,2\ra_{1,3}$ after the first beam splitter (BS$_1$).
Passing through the nonlinear sign gates, it
yields $-|2,0\ra_{1,3}+|0,2\ra_{1,3}$.
The second beam splitter (BS$_2$--conjugate to beam splitter 1)
then puts this into $-|1,1\ra_{1,3}$.
Obviously, all other input states, $|0\ra_L|1\ra_L$, $|0\ra_L|1\ra_L$,
$|1\ra_L|0\ra_L$, are not changed.
}
\end{center}
\end{figure}

The implementation of the desired operation is 
achieve by two 50:50 beam splitters and two nonlinear sign gates
(see Fig.\ 2), with probability of success 1/16.
Effectively then, a Kerr nonlinearity can be generated by
linear optics and projective measurements.
The probability of success then can be boosted by using gate-teleportation 
technique and sufficient number of ancilla photons.
It has been also demonstrated that such a nondeterministic two-qubit 
gate can be made for qubits defined by the 
polarization degree of freedom \cite{pittman01,pittman02a}.
A general formalism for the effective photon nonlinearities generated by
conditional measurement schemes in linear optics has been 
developed in some of our recent work \cite{lapaire03}.
Naturally, we emphasized that the ability to 
discriminate the number of incoming photons plays an essential role in the
realization of such nonlinear quantum gates in LOQC \cite{bartlett02,lee03,achilles03}.

\section{Optical lithography beyond diffraction limit}

Since the projective measurement can produce an
effective photon-photon
interaction, it can be a useful tool to
manipulate quantum correlations between photons.
A particularly interesting type of quantum state of light
is the maximally entangled photon-number state.
In our recent work, it has been shown that the Rayleigh diffraction limit in
optical lithography can be overcome \cite{boto00} by
using a quantum state of light of the following form:
\begarr
 {\textstyle \frac{1}{\sqrt{2}}}
\left( \left|N,0 \right\ra_{ab}
   + \left|0,N \right\ra_{ab}\right) ,
\label{noon}
\endarr

\no 
where $a,b$ denote two different paths.
It is well known that the $N=2$ path-entangled state of
Eq.~(\ref{noon}) can be generated using a Hong-Ou-Mandel
interferometer and two single-photon input states.
A 50:50 beam splitter, however, is not sufficient for producing
path-entangled states with a photon number larger than two \cite{campos89}.
On the other hand, the generation of these states 
with $N>2$ seems to involve a large Kerr 
nonlinearity, which
makes their physical implementation
very difficult \cite{gerry01}.
 
Using the technique of projective measurement, we have shown that 
by conditioning on single-photon--detection,
the generation of path-entangled photon-number states
is possible for more than two photons \cite{lee02a,kok02a}.
Figure 3 depicts a simple Mach-Zehnder type interferometric
scheme for producing such a state with $N=4$, using dual Fock-state
inputs, $|N\ra_a |N\ra_b$.

\begin{figure}[b]
   \begin{center}
   \begin{tabular}{c}
   \includegraphics[height=4cm]{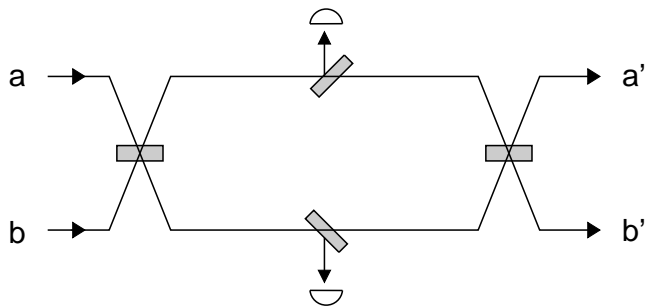}
   \end{tabular}
   \end{center}
   \caption[example] 
{
Mach-Zehnder interferometer with two additional beam splitters, 
which direct the reflected beams to photodetectors.
Conditioned on a specific outcome of photodetection,
a desired output state can be prepared in the mode $a'$ and $b'$.
}
\end{figure}

Suppose that we have the $|3,3\ra$ state as the input entering into
the modes $a$ and $b$.
Then, the first beam splitter transforms 
$|3,3\ra$ into a linear superposition of
$|6,0\ra$, $|4,2\ra$, $|2,4\ra$, and $|0,6\ra$.
After passing through the two intermediate beam splitters,
and if one and only one photon is counted at each detector,
the state is then projected onto an equal superposition of 
$|3,1\ra$ and $|1,3\ra$. 
Simply, the states $|6,0\ra$ or $|0,6\ra$ are discarded 
by this feedback from the photodetectors,
since they cannot yield a click at both detectors.
The $|4,2\ra$ and $|2,4\ra$ states, on the other hand,
lose one photon in each arm of the interferometer and are projected
to $|3,1\ra$ and $|1,3\ra$, respectively.
Thus, just before the last beam splitter, we have
a superposition of $|3,1\ra$ and $|1,3\ra$ with a known phase. 
We use an appropriate phase shifter in one of the two arms of
interferometer so that the state after the projective measurement
is reduced to $|3,1\ra - |1,3\ra$. 
Consequently after the last beam splitter, we get the desired state
$|4,0\ra - |0,4\ra$.
We have further shown that it is possible to produce
any two-mode, entangled, photon-number state
with only linear optical devices conditioned on photodetection \cite{kok02a}.
Although the probability of success generally decreases exponentially
as $N$ increases \cite{kok02a,fiurasek02,zou02,pryde03},
it was shown that
the scaling can be sub-exponential in $N$ by using quantum memory \cite{fiurasek03}. 
For some applications, however, it can already be
useful to have four-photon entanglement. 
Quantum interferometric lithography
is such an example.
Our approach has been used in a recent experiment to produce
maximally entangled three-photon 
polarization states \cite{mitchell03}.

\section{Phase-noise reduction beyond shot-noise limit}

In a typical optical interferometer, in which ordinary
coherent laser light enters via one input port, the phase sensitivity
in the shot-noise limit scales as $\Delta \varphi = 1/\sqrt{\bar N}$
where ${\bar N}$ is the mean number of photons.
Over the last two decades, a lot of effort was devoted to overcoming 
this limit, due to the obvious practical applications.
In the early 1980's, 
Caves first demonstrated that squeezing 
the vacuum noise in the unused
input port of an interferometer causes the  phase sensitivity to beat
the standard shot-noise limit
by $1/\sqrt{\bar N} \rightarrow 1/{\bar N}$
in the limit of infinite squeezing \cite{caves81}.
Bondurant and Shapiro proposed a multifrequency squeezed state interferometer
for this same purpose \cite{bondurant84}.
Hermann Haus pioneered in the generation of squeezed light
in optical fibers \cite{bergman91} 
as well as in Mach-Zehnder interferometers \cite{shirasaki94},
towards achieving the goal of Heisenberg-limited interferometry.

On the other hand, in 1986 it had been suggested by
Yurke and by Yuen  
that the phase-noise reduction can also be achieved using inputs with 
number eigenstates incident upon both input ports of a Mach-Zehnder
interferometer \cite{yurke86a,yuen86}.
In particular, Yurke and collaborators showed that if the photons
entered into each input port of the interferometer
in nearly equal numbers with a certain type of correlation, then,
it was possible to obtain an asymptotic phase
sensitivity of $1/N$, the Heisenberg limit \cite{yurke86b}.  
The so-called Yurke state is of the form:
\begarr
{\textstyle \frac{1}{\sqrt{2}} }
\left[
   \left|\mbox{${N}$},\mbox{${N-1}$} 
   \right\ra_{ab} +
   \left|\mbox{${N-1}$},\mbox{${N}$} \right\ra_{ab}
 \right],
\label{yurke}
\endarr

\no
where $a,b$ denote the two input modes.

Then, in the early 1990's, Holland and Burnett proposed 
Heisenberg-limited interferometry
by the use of so-called {\em dual Fock states} 
of the form 
$|N,N\ra_{ab}$
\cite{holland93}.
Such a
state can be approximately generated by degenerate parametric down
conversion or by optical parametric oscillation.
In a conventional Mach-Zehnder interferometer, 
only the difference of the number of photons at the
output is measured. 
However, to obtain increased sensitivity with dual Fock states, 
some special detection scheme is required, for which
Hall and co-workers proposed a combination of a direct measurement of the variance
of the difference current as well as a data-processing method based on
Bayesian analysis \cite{kim98}.
Other types of special input states have been proposed for
achieving the Heisenberg-limited 
phase sensitivity \cite{hillery93,brif96,berry00}.

\begin{figure}
   \begin{center}
   \begin{tabular}{c}
   \includegraphics[height=3.5cm]{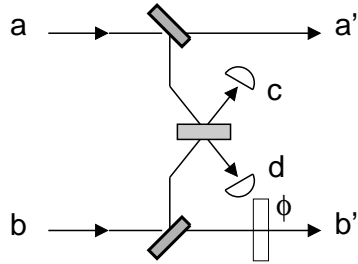}
   \end{tabular}
   \end{center}
   \caption[example] 
{A simple path-entanglement generator.
A Yurke-type quantum correlation between the two modes can be
produced with a dual Fock-state.
Suppose we post-select the outcome, conditioned upon
only one photon detection by either one of the two detectors.
Due to the 50:50 beam splitter in the midway, it is not possible to tell
whether mode $a$ or $b$ lost one photon.
The fundamental lack of which-path information
provides the entanglement between the two output modes.
For two-fold coincidence detection, the two detected photons
are from either mode $a$ or mode $b$, which eliminates the possibility of
peeling off one photon from each mode.
}
\end{figure}

In particular, the Yurke state approach has the same measurement scheme as
the conventional Mach-Zehnder interferometer;
a direct detection of the difference current \cite{dowling98}.
It is, however, not easy to generate the desired correlation
in the input state.
On the other hand, the dual Fock-state
approach finds a rather simple input state, but
requires a complicated data processing methods.
However, by a simple utilization of the projective measurements 
with linear optical devices,
it is possible to generate a desired
correlation in the Yurke state 
directly from the dual Fock state.

Let us consider a linear optical setup depicted in Fig.\ 4.
For a given dual Fock-state input $|N,N\ra_{ab}$, 
the output state conditioned on, for example, a
two-fold coincident count is given by
\begarr
 {\textstyle \frac{1}{\sqrt{2}}} \left[ |N,N-2\ra+ |N-2,N\ra \right] .
 \label{dowling}
\endarr
\no
It is not difficult to see that the condition of the coincident
detection yields either one of the two modes before the beam 
splitter must contain two photons while the other modes
contains no photon. 
This is an inverse-HOM situation where 
one photon at each mode cannot contribute 
to the coincident detection.
Consequently, 
the coincident detection results in a situation
where the main modes
$a$ and $b$ can only lose two photons or not at all.
Here the probability success of this event can be optimized by choosing
the reflection coefficient of the first beam splitters.
For the reflection coefficient of $|r|^2 = {1 / N}$,
its asymptotic value is found as $1/(2e^2)$, independent of $N$ \cite{lee02b}.
Furthermore, using a stack of such devices 
with appropriate phase shifters,
we have developed a method for the generation of 
maximally path-entangled states of the form 
Eq.\ (\ref{noon}) with an arbitrary number of photons \cite{kok02a}.

\section{Single-photon QND measurement devices}

In quantum optics the quantum nondemolition (QND)
devices are usually considered in the
context of photon-number measurements \cite{grangier98}.
In 1985 Imoto, Haus, and Yamamoto developed the basic idea of QND 
in quantum optics, which consists of
coupling the signal beam to the `meter' beam in a 
nonlinear medium and the detection of the phase shift 
of the meter beam measures the number of photons 
in the signal beam \cite{imoto85}.
The readouts of the number of photons in the signal beam
are performed by phase-sensitive homodyne detection 
of the meter beam in interferometer arrangements.

By the same token, as discussed in Section 1,
QND measurements at the single-photon level 
becomes extremely difficult due to the tiny
strength of the nonlinear interaction between photons.
In a recent experiment,
a single-photon QND has been demonstrated
by using a resonant coupling between a cavity field and the
meter atoms \cite{nogues99}.
Such a QND device at the single photon level
can provide a key tool for optical quantum information
processing, perhaps most importantly in quantum error correction. 
In contrast to the cavity approach,
we have proposed a probabilistic device
that signals the presence
of a single photon without destroying it 
using the technique of projective measurement \cite{kok02b}.

A simple way to perform a single-photon QND measurement is to use
quantum teleportation.
For example, a maximally polarization-entangled
photon pair produced by a parametric down-converter
can serve as a quantum channel.
If the input state is in a arbitrary superposition of 
zero and one photon with a fixed polarization,
the detector coincidence in Bell state measurement,
signals the present of a single
photon in the input and also the output states.
Simply, a vacuum input can never yield a two-fold detector coincidence.

This teleportation-based QND scheme works only if the input states
are restricted to one or zero photons. 
However, it breaks down if
there are more than two photons in the input.
For example, if the input state is of the form:
\begarr
|\psi\ra_{in}
&=& c_0 |0\ra + c_1 |1\ra +c_2 |2\ra \;,
\label{input}
\endarr

\no
the two-photon term will contribute to the two-fold coincidence 
even when the output of the down-converter is vacuum, yielding
a false identification of a single photon in the output state,
conditioned on a detector coincidence.

\begin{figure}
   \begin{center}
   \begin{tabular}{c}
   \includegraphics[height=3.5cm]{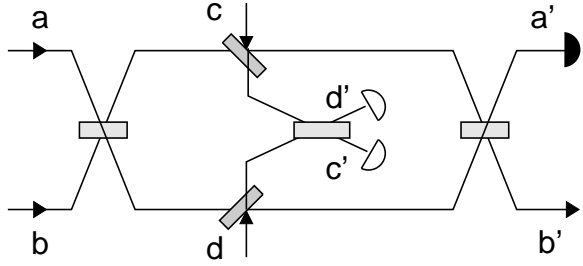}
   \end{tabular}
   \end{center}
   \caption[example] 
{QND measurement device for single-photon detection.
The input state, of an arbitrary superposition of $|0\ra$, $|1\ra$, and $|2\ra$,
enters into mode $a$, and
an auxiliary single photon is prepared for both modes $c$ and $d$.
Conditioned upon a detector coincidence in modes $c'$ and $d'$,
and no count in mode $a'$,
the outgoing mode $b'$ is a single-photon state.
}
\end{figure}

If we restrict the number of photons in the input up to two,
we can eliminate such a false identification by
using an interferometric setup depicted in Fig.5.
In Fig. 5, we assume that
the input state of the form Eq.(\ref{input}) enters into
in mode $a$,
and we further prepare single photons for mode $c$ and $d$.
Assuming beam splitters are 50:50,
the transformation of the probe photons in the mode $c$ and $d$
can then be written as
\begarr
\cd \dd &\rightarrow&
{\textstyle \frac{1}{4}}
(\hb^{\prime\dagger 2} -\ha^{\prime\dagger 2} 
+ \hd^{\prime\dagger 2} -\hc^{\prime\dagger 2}
-2 \ha^{\prime\dagger} \hc^{\prime\dagger}
+ 2 \hb^{\prime\dagger} \hd^{\prime\dagger} ).
\label{probe}
\endarr

\no     
Then we post select the photodetection outcome for
one and only one photon counted at each detector.
This condition requires either two photons are in mode $c$ or
two photons are in mode $d$, 
which eliminates the contribution from 
$c_0 |0\ra$ of the input state.
For one-photon and two-photon input states, we have
\begarr
\ha^{\dagger}
&\rightarrow&
( \ha^{\prime\dagger} - \hc^{\prime\dagger})/\sqrt{2} \;,
\qquad
\ha^{\dagger 2} 
\rightarrow
(\ha^{\prime\dagger 2} -2 \ha^{\prime\dagger}\hc^{\prime\dagger}
+\hc^{\prime\dagger 2}    
)/2 \; .
\label{two-photon}
\endarr

Now the only two-fold coincidence in the mode $c'$ and $d'$
by a two-photon input
is possible when the
$2 \hb^{\prime\dagger} \hd^{\prime\dagger}$ from Eq.(\ref{probe})   
and 
$ 2 \ha^{\prime\dagger} \hc^{\prime\dagger}$ from
Eq.(\ref{two-photon}) combine,
yielding 
$\ha^{\prime\dagger}\hb^{\prime\dagger}\hc^{\prime\dagger}
\hd^{\prime\dagger}$.
However, further postselecting on the vacuum in the mode $a'$
eliminates this two-photon contribution to
the two-fold coincidence in $c'$ and $d'$.

A single photon in mode $a$ yields a contribution
$\hb^{\prime\dagger} \hc^{\prime\dagger} \hd^{\prime\dagger}$,
indicating that there is a two-fold coincidence in mode $c'$ and $d'$, and
a single photon in the output mode $b'$.
As can be seen if Eqs.(\ref{probe},\ref{two-photon}),
the probability of success for this interferometric device is given
by 1/8.
By adjusting the transmission coefficients of the beam splitters in
modes $c$ and $d$, the probability of success
can be increased further.
Obviously, this scheme does not work when the incoming state has
an unknown polarization.
However, it turns out that a more sophisticated interferometric setup
with polarization beam splitters
can do the job 
while preserving the unknown polarization \cite{kok02b}.
Of course, such a scheme is not a full QND measurement of 
the photon-number observable, 
since it works for only zero, one, and two photons.
It can, however, still play an important role in linear optical
quantum computation, where up to only two photons are used in 
each logic gate \cite{pittman03,obrien03}.
Furthermore, such a single-photon QND device can be used
in various quantum communication protocols
such as quantum repeaters \cite{jacobs02,kok03}.

\section{Summary}

Linear optics with projective measurements,
can be used to replace the use Kerr nonlinearities
and provide a much higher efficiency.
Using this technique, we have studied the generation of useful
photonic quantum correlations.
The maximally path-entangled photon-number states
provide an essential way for optical lithography to
proceed beyond the Rayleigh diffraction limit.
The Yurke-type path-entanglement is of particular 
importance in Heisenberg-limited interferometry. 
Projective measurements also enable us to 
construct a device that signals the presence of a single photon 
without destroying it.
Single-photon non-demolition measurement 
is of great importance in quantum information
processing with photons,
since most error-correction codes in the presence of 
qubit loss requires QND measurements \cite{gingrich03}.

\ack

This work was carried out at the Jet Propulsion Laboratory, California
Institute of Technology, under a contract with the National Aeronautics 
and Space Administration. 
The authors wish to thank
C.\ Adami, N.J.\ Cerf, J.D.\ Franson, 
G.J.\ Milburn, W.J.\ Munro, T.B.\ Pittman, and T.C.\ Ralph 
for helpful discussions. 
We would like to acknowledge support from 
the NASA Intelligent Systems Program,
the National Security Agency,
the Advanced Research and Development Activity, 
the Defense Advanced Research Projects Agency,
the National Reconnaissance Office, the Office of Naval Research,
and the Army Research Office.
P.K.\ thanks the National Research Council and NASA Code Y
for support.


\section*{References}

\begin{thebibliography}{99}

\bibitem{boyd91}
Boyd RW 1991
{\it Nonlinear Optics} (Academic Press, San Diego, CA)


\bibitem{klm01}
Knill E, Laflamme R, and Milburn GJ 2001
{\it Nature} {\bf 409} 46


\bibitem{scully97}
Scully MO and Zubairy MS 1997
{\it Quantum Optics}
(Cambridge University Press, Cambridge, UK)

\bibitem{turchette95}
Turchette QA, Hood CJ, Lange W, Mabuchi H, and Kimble HJ 1995
{\it Phys. Rev. Lett.} {\bf 75} 4710

\bibitem{milburn89}
Milburn GJ 1989 
{\it Phys. Rev. Lett.} {\bf 62} 2124 

\bibitem{ralph01}
Ralph TC, White AG, Munro WJ, and Milburn GJ 2001 
{\it Phys. Rev. A} {\bf 65} 012314

\bibitem{hom87}
Hong CK, Ou ZY, and Mandel L 1987
{\it Phys. Rev. Lett.} {\bf 59} 2044


\bibitem{pittman01}
Pittman TB, Jacobs BC, and Franson JD 2001
{\it Phys. Rev. A} {\bf 64} 062311

\bibitem{pittman02a}
Pittman TB, Jacobs BC, and Franson JD 2002
{\it Phys. Rev. Lett.} {\bf 88} 257902

\bibitem{lapaire03}
Lapaire GG, Kok P, Dowling JP, and Sipe JE 2003
{\it Phys. Rev. A} {\bf 68} 042314


\bibitem{bartlett02}
Bartlett SD, Diamanti E, Sanders BC, and Yamamoto Y 2002
{\it Proceedings of Free-Space Laser Communication and Laser Imaging II}
{Vol. 4821} (SPIE, Bellingham, WA)
quant-ph/0204073


\bibitem{lee03}
Lee H, {\em et al}. 2003
``Towards photostatistics from photon-number discriminating detectors,''
(unpublished) quant-ph/0310161


\bibitem{achilles03}
Achilles D, {\em et al}. 2003
``Photon number resolving detection using time-multiplexing,''
(unpublished) quant-ph/0310183


\bibitem{boto00}
Boto AN, {\em et al.} 2000
{\it Phys. Rev. Lett.} {\bf 85} 2733


\bibitem{campos89} 
Campos RA, Saleh BEA, and Teich MC 1989
{\it Phys. Rev. A} {\bf 40} 1371

\bibitem{gerry01}
Gerry CC and Campos RA 2001
{\it Phys. Rev. A} {\bf 64} 063814

\bibitem{lee02a}
Lee H, Kok P, Cerf NJ, and Dowling JP 2002
{\it Phys. Rev. A} {\bf 65} 030101

\bibitem{kok02a}
Kok P, Lee H, and Dowling JP 2002
{\it Phys. Rev. A} {\bf 65} 052104

\bibitem{fiurasek02}
Fiur\'asek J 2002
{\it Phys. Rev. A} {\bf 65} 053818

\bibitem{zou02}
Zou XB, Pahlke K, and Mathis W 2002
{\it Phys. Rev. A} {\bf 66} 014102

\bibitem{pryde03}
Pryde GJ and White AG 2003
{\it Phys. Rev. A} {\bf 68} 052315

\bibitem{fiurasek03}
Fiur\'asek J, Massar S, and Cerf NJ 2003
{\it Phys. Rev. A} {\bf 68} 042325

\bibitem{mitchell03}
Mitchell MW, Lundeen JS, Steinberg AM 2003
``Super-resolving phase measurements with a multi-photon
entangled state,'' (unpublished)
quant-ph/0312186


\bibitem{caves81}
Caves CM 1981
{\it Phys. Rev. D} {\bf 23} 1693

\bibitem{bondurant84}
Bondurant RS and Shapiro JH 1984
{\it Phys. Rev. D} {\bf 30} 2548

\bibitem{bergman91}
Bergman K and Haus HA 1991
{\it Opt. Lett.} {\bf 16} 663




\bibitem{shirasaki94}
Shirasaki M, Lyubomirsky, and Haus HA 1994
{\it J. Opt. Soc. Am. B} {\bf 11} 857


\bibitem{yurke86a}
Yurke B 1986
{\it Phys. Rev. Lett} {\bf 56} 1515


\bibitem{yuen86}
Yuen HP 1986
{\it Phys. Rev. Lett.} {\bf 56} 2176

\bibitem{yurke86b}
Yurke B, McCall SL, and Klauder JR 1986
{\it Phys. Rev. A} {\bf 33} 4033


\bibitem{holland93}
Holland MJ and Burnett K 1993
{\it Phys. Rev. Lett.} {\bf 71} 1355 


\bibitem{kim98} 
Kim K, 
Pfister O, Holland MJ, Noh J, Hall JL 1998
{\it Phys. Rev. A} {\bf 57} 4004

\bibitem{hillery93}
Hillery M and Mlodinow L 1993
{\it Phys. Rev. A} {\bf 48} 1548

\bibitem{brif96}
Brif C and Mann A 1996
{\it Phys. Rev. A} {\bf 54} 4505

\bibitem{berry00}
Berry DW and Wiseman HM 2000
{\it Phys. Rev. Lett.} {\bf 85} 5098


\bibitem{dowling98}
Dowling JP 1998
{\it Phys. Rev. A} {\bf 57} 4736

\bibitem{lee02b}
Lee H, Kok P, and Dowling JP 2002 
{\it J. Mod. Opt.} {\bf 49} 2325

\bibitem{grangier98}
Grangier P, Levenson JA, and Poizat PJ 1998
{\it Nature} {\bf 396} 537

\bibitem{imoto85}
Imoto N, Haus HA, and Yamamoto Y 1985
{\it Phys. Rev. A} {\bf 32} 2287


\bibitem{nogues99}
Nogues G, {\it et al}. 1999
{\it Nature} {\bf 400} 239


\bibitem{kok02b} 
Kok P, Lee H, and Dowling JP 2002
{\it Phys. Rev. A} {\bf 66} 063814


\bibitem{pittman03}
Pittman TB, Fitch MJ, Jacobs BC, and Franson JD 2003
{\it Phys. Rev. A} {\bf 68} 032316



\bibitem{obrien03}
O'Brien JL, Pryde GJ, White AG, Ralph TC, and Branning D 2003
{\it Nature} {\bf 426} 264

\bibitem{jacobs02}
Jacobs BC, Pittman TB, and Franson JD 2002
{\it Phys. Rev. A} {\bf 66} 052307


\bibitem{kok03}
Kok P, Williams CP, and Dowling JP 2003
{\it Phys. Rev. A} {\bf 68} 022301



\bibitem{gingrich03}
Gingrich RM, Kok P, Lee H, Vatan F, and Dowling JP 2003
{\it Phys. Rev. Lett.} {\bf 91} 217901



\end{thebibliography}
\end{document}